\documentstyle[12pt]{article}
\newcommand{\bm}[1]{ \mbox{\boldmath $ #1 $} }
\newcommand{\eq}{\,=\,}
\newcommand{\abs}[1]{\mid #1 \mid}
\newcommand{\uln}[1]{\underline{#1}}
\newcommand{\Snc}[2]{S(#1,a)(#2)}
\newcommand{\Sn}[1]{
{\sin{\left[{\pi\over a} \left(x\,-\, #1 a\right)\right]}\over
{\pi\over a}\left(x\,-\, #1 a\right)}
}

\begin{document}

\title{
{\bf Solving the Coulomb Schr\"{o}dinger Equation in d = 2+1 via Sinc
Collocation\thanks{
 preprint UTAH-IDR-CP-05}
}
      }
\author{
        Vasilios G. Koures\thanks{Email: koures@dirac.chem.utah.edu}\\
{\em Department of Chemistry, Henry Eyring Bldg., Univ. of Utah\/}\\
{\em Salt Lake City, UT 84112 \ \ USA\/}\\
       }
\date{}
\vspace{0.3cm}
\maketitle
\setlength{\baselineskip}{4.0ex}

\begin{abstract}
\setlength{\baselineskip}{4.0ex}
We solve the non-relativistic Coulomb Shr\"{o}dinger equation in
d = 2+1 via sinc collocation. We get excellent convergence using
a generalized sinc basis set in position space. Since convergence
in position space could not be obtained with more common numerical
techniques, this result helps to corroborate the conjecture that
the use of a localized basis set within the context of light cone
quantization can yield much better convergence. All of the
computations presented here were performed on an IBM-compatible
PC with an Intel 486DX2-66 microchip.
\end{abstract}

	Recently, light cone quantization (LCQ) of quantum field theory has attracted
considerable attention as a possible alternative method for solving
non-perturbative
problems in quantum field theory\cite{brod}. In this scheme, one obtains  a
rational,
closed-form, and relativistically covariant Hamiltonian which allows one to
avoid many
of the severe mathematical difficulties which have plagued traditional
equal time quantization techniques\cite{brod}. The discrete version of the
light
cone Hamiltonian allows one to model gauge theory on a computer as
an eigenvalue and matrix diagonalization problem over a discrete and
covariantly regularized Fock space\cite{tang}.

A numerical implementation of the Discretized Light Cone Quantization (DLCQ)
method to study positronium has produced promising results, but it is evident
that the use of a plane wave basis will be a major bottleneck
when one applies this method to more complicated models\cite{paul}. It has been
suggested that the use of a localized basis set would yield much better
convergence for bound state computations\cite{wilson,kour}.
The method introduced in Ref. [5] has now been adapted
to the orthogonal localized basis set of Sinc functions.

Sinc methods have increasingly been recognized
as very powerful tools for attacking problems within
applied physics and engineering\cite{stenger}. Until now, however,
it was not clear how to apply Sinc methods within the context of the
operator formalism of quantum field theory. This general formalism will be
reported in detail elsewhere; the focus of the present
paper is to demonstrate the power of the sinc collocation method by
solving the radial Coulomb equation in d = 2+1.

The motives for studying quantum electrodynamics (QED) in 2+1 dimensions are
numerous\cite{tam}. The lower dimensions allow a smaller number of degrees of
freedom but the model still possesses independent photon degrees of freedom,
unlike the (1+1)-dimensional model. The model is super-renormalizable and,
when formulated with four-component spinors, it exhibits
confinement\cite{tam,yung}.

The non-relativistic Coulomb Schr\"{o}dinger equation is derived from the LCQ
formalism as follows\cite{tam}. One first derives the discretized light-cone
Hamiltonian for (2+1)-dimensional QED with four-component spinors. A
Tamm-Dancoff integral equation is then obtained for the ``positronium''
bound states. Taking the weak-coupling limit, one gets a non-relativistic
integral equation which is the momentum-space Coulomb Schr\"{o}dinger
equation; the infrared divergences cancel between the self-mass and one-photon
exchange diagrams\cite{tam}. The position-space result is then obtained by a
Fourier transform,
\begin{equation}
\left[-\,{1\over m}\bigtriangledown_{r}^2\,+\,{g^2\over 2\pi}\left(
\gamma\,+\,\ln mr\right)\right]\Psi(\bm{r})\eq E\,\Psi(\bm{r}) \qquad,
\label{coul}
\end{equation}
where $\gamma$ is the Euler-Mascheroni constant, $m$ is the mass, and
$g$ is the coupling constant. It is convenient to rewrite eq.~(\ref{coul})
in terms of the dimensionless variables
$\bm{x} \eq \sqrt{{mg^2\over 2\pi}}\,\bm{r}$ and
$\lambda' \eq {2\pi\over g^2}\,E\,+\,\ln \sqrt{{2g^2\over m\pi}}$. We get
\begin{equation}
\left(-\bigtriangledown_{x}^2\,+\,\ln x \,+\,\gamma\,+\,\ln 2\right)
\Psi(\bm{x}) \eq \lambda'\,\Psi(\bm{x}) \qquad, \label{coul1}
\end{equation}
or
\begin{equation}
\left(-\bigtriangledown_{x}^2\,+\,\ln x\right)\Psi(\bm{x})
\eq \lambda\,\Psi(\bm{x}) \qquad, \label{coul2}
\end{equation}
with
\begin{equation}
\lambda \eq \lambda'\,-\,\gamma\,-\,\ln 2 \qquad. \label{lamparam}
\end{equation}

The separation of variables,
$\Psi(\bm{x})\eq R(x)\Theta(\theta)$, leads to
$\Theta(\theta)\eq\exp(\pm il\theta)$, where $l$ is the angular momentum
quantum number. We are left with a radial differential equation for $R$
which, after using the substitution\cite{yung},
\begin{equation}
R(x)\eq x^{-1/2}\,f(x)  \qquad,  \label{sub}
\end{equation}
transforms into a differential equation for $f$:
\begin{equation}
-f''(x)\,+\,\left[{4l^2\,-\,1\over 4x^2}\,+\,\ln x\right]\,f(x)
\eq \lambda\,f(x)  \qquad. \label{ode}
\end{equation}
This equation for $f$ represents a singular Sturm-Liouville
system and it can be solved by sinc collocation\cite{stenger}.

To this end, we begin with the definition of the Sinc function.
If $a > 0$ and $m$ is an integer, the {\em Sinc\/} function,
$S(m,a)(x)$, is defined by
\begin{equation}
S(m,a)(x)\,=\, \Sn{m}  \ . \label{sncdf}
\end{equation}
The theory of sinc series (cardinal functions) on the entire real line has
been thoroughly developed. For a class of functions known as the Payley-Weiner
class, the sinc interpolation and quadrature formulas are exact\cite{stenger}.
However, a more practical application of sinc approximation of functions which
are in a much less restrictive class has also been developed and the absolute
errors have been derived via contour integration\cite{stenger}.
In a nutshell, the Paley-Wiener functions are entire but a more practical
class of functions should have specific growth restrictions on the real
line and should be analytic only on an infinite strip centered about
the real line,
\begin{equation}
D_S\,\equiv\, \left\{ z\,\epsilon\,{\bm{\cal C}}\,:\,
z\eq x\,+\,iy\ ,\ \abs{y}\,<\,d\right\} \qquad. \label{doms}
\end{equation}
$\bm{\cal C}$ denotes the set of complex numbers. It turns out that the
absolute
error of sinc interpolation and quadrature on such functions is
exponentially damped\cite{stenger}.

For the problem of interest to us we need to use sinc methods on a function
$f$ whose domain is $(0,\infty)$. The more general class just described,
however,
has a domain which includes the whole real line. This conflict is elegantly
circumvented through the use of conformal maps\cite{stenger}. For example,
let $\phi$ be a one-to-one conformal map from some domain $D$ to domain $D_S$
and let $\psi$ denote the inverse map which is also conformal. If $f$ is
analytic in $D$ then $f\circ\psi$ is analytic in $D_S$. So, if a
numerical process has been developed in a domain containing the whole real
line, $\bm{\cal R}$, then this process can be carried over to a new domain
containing only a proper subset of the real line.

We will shortly see that the domain of interest to us is
\begin{equation}
D\,\equiv\,
\left\{w\,\epsilon\,{\bm{\cal C}}\,:\,
\abs{\arg(sinh(w))}\,<\,d\,\leq\,{\pi\over2}\right\}
\qquad. \label{dom}
\end{equation}
This domain is conformally mapped onto the infinite strip $D_S$ by the
function\cite{stenger}
\begin{equation}
z\eq \phi(w)\eq \ln(sinh(w)) \qquad. \label{phidf}
\end{equation}
If we let
\begin{equation}
w\eq \psi(z) \eq\phi^{-1}(z)\eq\ln(\,e^z\,+\,\sqrt{1\,+\,e^{2z}}\,)
\qquad, \label{psidf}
\end{equation}
then
\begin{equation}
\Gamma\,\equiv\, \psi({\bm{\cal R}})\eq (0,\infty) \qquad, \label{Gammadf}
\end{equation}
as desired. For $a\,>\,0$ and $m$ an integer we define the {\em sinc points \/}
\begin{equation}
x_m\,\equiv\,
\psi(ma)\eq\ln\left(\,e^{ma}\,+\,\sqrt{\,1\,+\,e^{2ma}\,}\,\right)
\qquad.   \label{sincpts}
\end{equation}
One may further verify that
\begin{eqnarray}
\phi'(x_m)&=& \sqrt{\,1\,+\,e^{-2ma}\,} \nonumber \\
\phi''(x_m)&=& -e^{-2ma} \qquad, \label{deriv}
\end{eqnarray}
where the primes denote differentiation with respect to x.

Given the above definitions and maps, suppose we have a function $f$ and
that there are positive constants $\alpha$, $\beta$, and $C$ such that
\begin{equation}
\abs{f(x)}\,\leq\,C
\left\{
  \begin{array}{ll}
  x^\alpha\ ,&\  x\,\epsilon\,\left(0,\ln(\,1+\sqrt{2}\,)\right) \\
  e^{-\beta x}\ ,&\  x\,\epsilon\,\left[\ln(\,1+\sqrt{2}\,),\infty\right)
  \end{array}
\right. \qquad. \label{asyma}
\end{equation}
If we choose
\begin{eqnarray}
N&=&ceil({\alpha\over\beta}\,M) \nonumber \\
a&=&\sqrt{\,{2\pi d\over\alpha M}\,}\,\leq\, {2\pi d\over\ln(2)}\ ,
\label{params}
\end{eqnarray}
where $ceil(x)$ rounds $x$ to the nearest integer $\geq x$, then\cite{stenger}
\begin{equation}
\abs{ \,f(x)\,-\,\sum_{m=-M}^{N}\,f(x_m)\,\Snc{m}{\phi(x)}\, }\eq
{\cal O}\left(\,\exp -\sqrt{\,\pi d\alpha M\,}\,\right) \ , \label{conv1}
\end{equation}
and
\begin{equation}
\abs{\,{d^k f\over dx^k}\,-\,{d^k\over dx^k}\sum_{m=-M}^N f(x_m)\,
\Snc{m}{\phi(x)}\,}\eq
{\cal O}\left(\,\exp -\sqrt{\,\pi d\alpha M\,}\,\right) \ . \label{conv2}
\end{equation}
As promised, we see that the absolute errors are exponentially damped.

The two equations (\ref{conv1}) and (\ref{conv2}) are all we need to solve the
differential equation (\ref{ode}). In addition, we would like to have a
normalized
final answer and so we also introduce the result for sinc
quadrature\cite{stenger}.
Suppose we have a function $F$ and that there are positive constants
$\alpha$, $\beta$, and $C$ such that
\begin{equation}
\abs{F(x)}\,\leq\,C
\left\{
  \begin{array}{ll}
  x^{\alpha-1}\ ,&\  x\,\epsilon\,\left(0,\ln(\,1+\sqrt{2}\,)\right) \\
  e^{-\beta x}\ ,&\  x\,\epsilon\,\left[\ln(\,1+\sqrt{2}\,),\infty\right)
  \end{array}
\right. \qquad. \label{asymb}
\end{equation}
Then\cite{stenger}
\begin{equation}
\abs{\,\int_{0}^{\infty}\!F(x)\,dx\,-\,
a\sum_{m=-M}^{N}{F(x_m)\over \phi'(x_m)}\,}\eq
{\cal O}\left(\,\exp -\sqrt{\,2\pi d\alpha M\,}\,\right) \ , \label{conv3}
\end{equation}
with $M$, $N$, and $a$ being defined the same as before.

Now, to solve (\ref{ode}) we note that as $x\rightarrow 0$ the physically
acceptable solution takes the form\cite{yung}
\begin{equation}
f\,\propto\,x^{l+1/2} \qquad. \label{zeroasym}
\end{equation}
{}From (\ref{asyma}) we may thus take $\alpha\eq l\,+\,1/2$.
On the other hand, as $x\rightarrow\infty$ the physically acceptable solution
takes the form
\begin{equation}
f\,\propto\,e^{-x\ln x} \qquad. \label{inftasym}
\end{equation}
So, from (\ref{asyma}) we may take $\beta\leq 1$. Next, we use (\ref{conv1})
and (\ref{conv2}) to approximate $f$ and $f''$, respectively:
\begin{eqnarray}
f(x)&\approx&\sum_{m=-M}^{N}\,\Snc{m}{\phi(x)}\,f(x_m) \label{approxa} \\
f''(x)&\approx&\!\!\sum_{m=-M}^{N}\!\left\{\left[(\phi'(x))^2\,
{d^2\over d\phi^2}\,\,+\,\phi''(x)\,
{d\over d\phi}\right]\!\Snc{m}{\phi(x)}\right\}\!f(x_m). \label{approxb}
\end{eqnarray}
To evaluate these expansions at a general sinc point $x_n$ note that
\begin{eqnarray}
\Snc{m}{\phi(x)}\, |_{x_n} &=& \delta^{(0)}_{n,m} \nonumber \\
{d\over d\phi}\,\Snc{m}{\phi(x)}\, |_{x_n} &=& {1\over a}\delta^{(1)}_{n,m}
\nonumber \\
{d^2\over d\phi^2}\,\Snc{m}{\phi(x)}\, |_{x_n} &=& {1\over
a^2}\delta^{(2)}_{n,m}
\qquad, \label{toeplitz}
\end{eqnarray}
where $\delta^{(0)}_{n,m}$ is the Kronecker delta function and
\begin{equation}
\delta^{(1)}_{n,m}\eq
\left\{
  \begin{array}{ll}
  0\ ,&\  m=n \\
  {(-1)^{m-n}\over m-n}\ ,&\  m\neq n
  \end{array}
\right. \label{delt1}
\end{equation}
\begin{equation}
\delta^{(2)}_{n,m}\eq
\left\{
  \begin{array}{ll}
  -\,{\pi^2\over 3}\ ,&\  m=n \\
  {2(-1)^{m-n+1}\over (m-n)^2}\ ,&\  m\neq n
  \end{array}
\right. \qquad. \label{delt2}
\end{equation}
Using these results along with (\ref{deriv}) we may now approximate (\ref{ode})
via sinc collocation,
\begin{equation}
\sum_{m=-M}^{N}\left[I^{(0)}_{n,m}\,+\,I^{(1)}_{n,m}\,+\,I^{(2)}_{n,m}\right]
f(x_m)\eq\lambda\,f(x_n) \ , \label{colloc}
\end{equation}
where
\begin{eqnarray}
I^{(0)}_{n,m}&=&\left({4l^2\,-\,1\over 4x^{2}_{m}}\,+\,\ln
x_m\right)\delta^{(0)}_{n,m}
\nonumber \\
I^{(1)}_{n,m}&=&{1\over a}\,e^{-2ma}\ \delta^{(1)}_{n,m} \nonumber \\
I^{(2)}_{n,m}&=&-\,{1\over a^2}(1\,+\,e^{-2ma})\ \delta^{(2)}_{n,m} \ .
\label{Idef}
\end{eqnarray}
In the tradition of LCQ, we have formulated the problem as an eigenvalue and
matrix diagonalization problem. The components of the eigenvector $f(x_m)$ can
be substituted into (\ref{approxa}) to compute the eigenfunction at arbitrary
$x$.
Finally, we use (\ref{sub}) to acquire $R(x)$ and we normalize this function by
computing the norm, $\sqrt{\,\int_{0}^{\infty}R^2(x)\,dx\,}$, using the sinc
quadrature formula (\ref{conv3}).

In Table 1 we list the first five eigenvalues for $l$ ranging from 0 to 4. The
diagonalization was performed with MATLAB\cite{matlab}. We used $d=\pi/4$,
$\beta$ = 1/2 to 1, and values for $M$ up to 500.
\begin{table}[htb]
\caption{ First five eigenvalues for $l$ ranging from 0 to 4.}
\begin{center}
\begin{tabular}{|c|c|c|c|c|c|} \hline
\ & $l\eq 0$ & $l\eq 1$ & $l\eq 2$ & $l\eq 3$ & $l\eq 4$  \\ \hline
$\lambda_{0}$&0.52643626&1.3861862&1.8443720&2.1578468&2.3962798\\ \hline
$\lambda_{1}$&1.6619365 &2.0094748&2.2758614&2.4881158&2.6638815\\ \hline
$\lambda_{2}$&2.1870578 &2.3943387&2.5800522&2.7390550&2.8772701\\ \hline
$\lambda_{3}$&2.5153639 &2.6726676&2.8144703&2.9409664&3.0543788\\ \hline
$\lambda_{4}$&2.7677810 &2.8906069&3.0049630&3.1096821&3.3373990\\ \hline
\end{tabular}
\end{center}
\end{table}
The convergence was excellent for
all $l$ values but we needed considerably higher $M$ values for $l=0$ than for
$l\neq 0$. In Ref.~[7] the first five $l=0$ eigenvalues were computed in
momentum
space:
\begin{eqnarray}
\lambda'_{0}\eq 1.7969\ ,&\qquad\lambda'_{1}\eq 2.9316\ ,\qquad\lambda'_{2}\eq
3.4475
\nonumber \\
\lambda'_{3}\eq 3.7858\ ,&\qquad\lambda'_{4}\eq 4.0380 \qquad, \label{tameigen}
\end{eqnarray}
where $\lambda'$ is related to $\lambda$ in Table 1 by (\ref{lamparam}). These
momentum space results do not agree with the position space
computations\cite{yung}
and additional repeated attempts to get convergence in position space have
failed\cite{tam}. It was concluded in Ref.~[7] that ``the previous position
space
calculation\cite{yung} was inaccurate, due to the slow, logarithmic behavior of
the
potential $\ldots$ The momentum space calculation is much more rapidly
convergent''.

On the other hand, if we use (\ref{lamparam}) to convert the
first column of Table 1 to the parameterization used in Ref.~[7] we see that
\begin{eqnarray}
\lambda'_{0}\eq 1.7967991\ ,&\qquad\lambda'_{1}\eq 2.9322993\ ,
\qquad\lambda'_{2}\eq 3.4574206  \nonumber \\
\lambda'_{3}\eq 3.7857268 \ ,&\qquad\lambda'_{4}\eq 4.0381439 \qquad;
\label{myeigen}
\end{eqnarray}
These values are in very good agreement with (\ref{tameigen}), they converge
faster, and they are more accurate.
Note further that the eigenvalues for $l\neq 0$ are actually in good
agreement with the position space results\cite{yung}. (The momentum space
calculations in [7] were not carried out for $l\neq 0$.) This leads us to
conclude that the slow convergence of the previous position space results
was not due to the slow logarithmic behavior of the potential; rather,
it was due to a minor instability caused by the sign-flip of the
``centrifugal''
part of the potential when $l=0$.

As promised, it is now very easy to use (\ref{approxa}) along with
(\ref{sub}) to compute any eigenfunction; these eigenfunctions are easily
normalized through the use of the sinc quadrature result (\ref{conv3}).
As a representative sample, in Figures (1) to (3) we display the first three
normalized
eigenfunctions for $l=0$; note that the number of bumps increases incrementally
as we go from $n=0$ to $n=2$. Figures (4) to (6) show the first three
normalized eigenfunctions for $l=4$; note that the ``centrifugal'' barrier
causes the bumps to move away from the center, relative to the $l=0$ plots.

We have thus demonstrated that sinc methods provide a very powerful tool for
solving the radial Coulomb Schr\"{o}dinger equation in $d=2+1$. These methods
are very accurate and converge very fast. This point is strongly demonstrated
by the fact that convergence for the $l=0$ values could not be attained in
position space by other well-known numerical techniques. The sinc collocation
method used here may be straightforwardly extended to attack more complicated
problems within quantum field theory; this formalism will be reported in
detail in the future.

\leftline{\uln{Acknowledgements}}
It is a pleasure to thank professor Frank Stenger for his many valuable
consultations. I also thank professor Frank Harris for his continued
encouragement and for stimulating conversations and suggestions.
This work was supported in part by the Office of Naval Research, contract
N0014-93-1-0196.

\bibliographystyle{unsrt}

\begin{thebibliography}{99.}

\bibitem{brod} S.J. Brodsky, G. McCartor, H.-C. Pauli, and S. Pinsky,
{\em Particle World\/} {\bf 3} (1993) 109.
\bibitem{tang} A.C. Tang, S.J. Brodsky, and H.-C. Pauli, {\em Phys. Rev.\/}
{\bf D44} (1991) 1842; A.C. Tang, SLAC report SLAC-351 (1991).
\bibitem{paul} M. Kalu\u{z}a and H.-C. Pauli, {\em Phys. Rev.\/}
{\bf D45} (1992) 2968; M. Krautg\"{a}rtner, H.-C. Pauli, and
F. W\"{o}lz, {\em Phys. Rev.\/} {\bf D45} (1992) 3755.
\bibitem{wilson} A. Harindranath, R.J. Perry, and J. Shigemitsu, {\em Phys.
Rev.\/}
{\bf D46} (1992) 4580.
\bibitem{kour} V.G. Koures, {\em Phys. Lett.\/} {\bf B348} (1995) 170.
\bibitem{stenger} F. Stenger,``Numerical Methods Based on Sinc and Analytic
Functions'', Springer-Verlag, NY, 1993.
\bibitem{tam} A. Tam, C.J. Hamer, and C.M. Yung, {\em Gen. Phys.\/} {\bf G},
in press; New South Wales University Preprint PRINT-94-0182 (1994).
\bibitem{yung} C.M. Yung and C.J. Hamer, {\em Phys. Rev.\/} {\bf D44} (1991)
2595.
\bibitem{matlab} John Little and Cleve Moller,``MATLAB User's Guide'', The
Mathworks, Inc., Natick, MA, 1993.

\end{thebibliography}

\begin{figure}
   \vspace{7.5in}
   \caption{Plot of normalized function $R_{0,0}(x)$ for $l\eq 0$
    and $\lambda_{0}\eq 0.5264$.}
\end{figure}
\begin{figure}
   \vspace{7.5in}
   \caption{Plot of normalized function $R_{1,0}(x)$ for $l\eq 0$
    and $\lambda_{1}\eq 1.6619$.}
\end{figure}
\begin{figure}
   \vspace{7.5in}
   \caption{Plot of normalized function $R_{2,0}(x)$ for $l\eq 0$
    and $\lambda_{2}\eq 2.1871$.}
\end{figure}
\begin{figure}
   \vspace{7.5in}
   \caption{Plot of normalized function $R_{0,4}(x)$ for $l\eq 4$
    and $\lambda_{0}\eq 2.3963$.}
\end{figure}
\begin{figure}
   \vspace{7.5in}
   \caption{Plot of normalized function $R_{1,4}(x)$ for $l\eq 4$
    and $\lambda_{1}\eq 2.6639$.}
\end{figure}
\begin{figure}
   \vspace{7.5in}
   \caption{Plot of normalized function $R_{2,4}(x)$ for $l\eq 4$
    and $\lambda_{2}\eq 2.8773$.}
\end{figure}

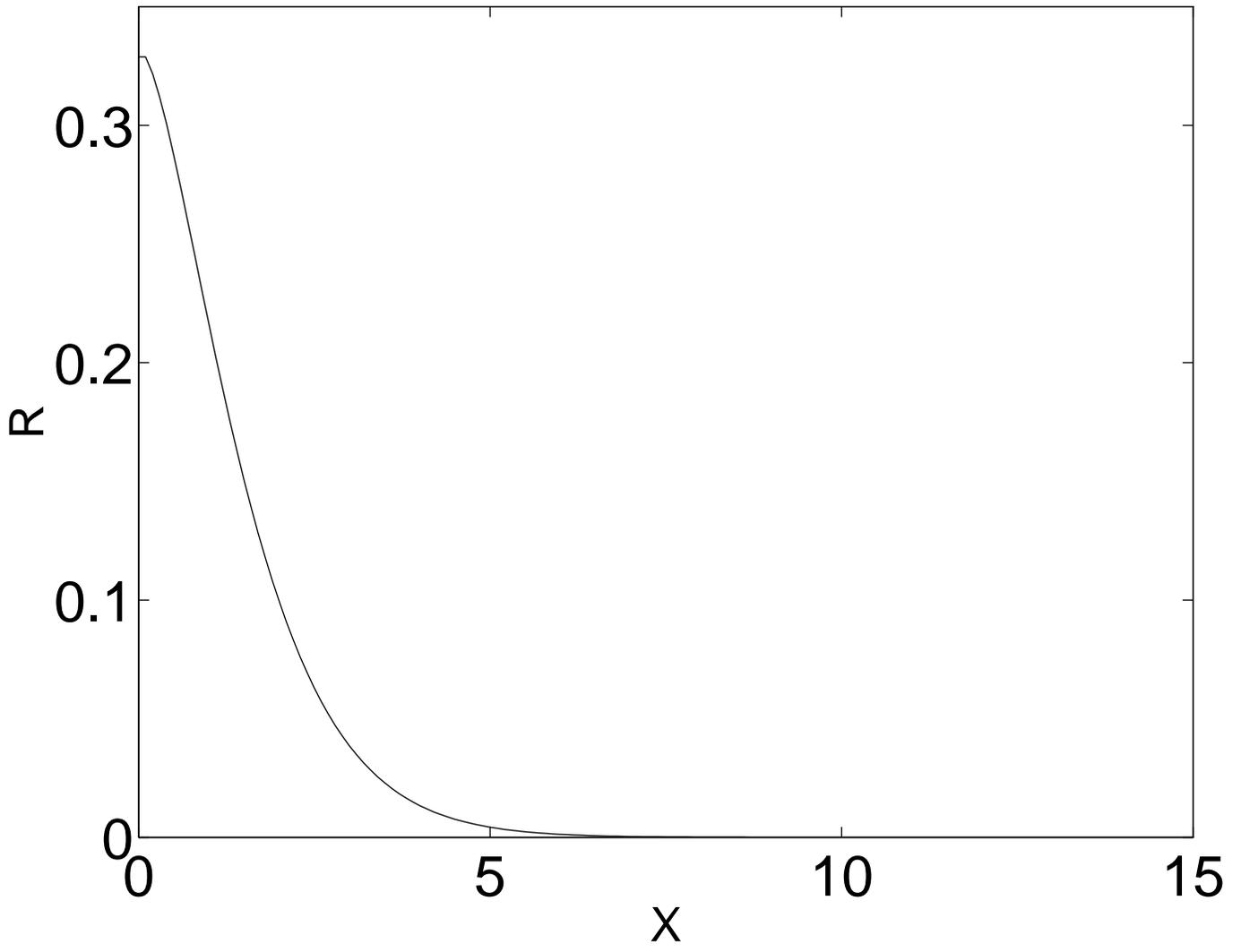
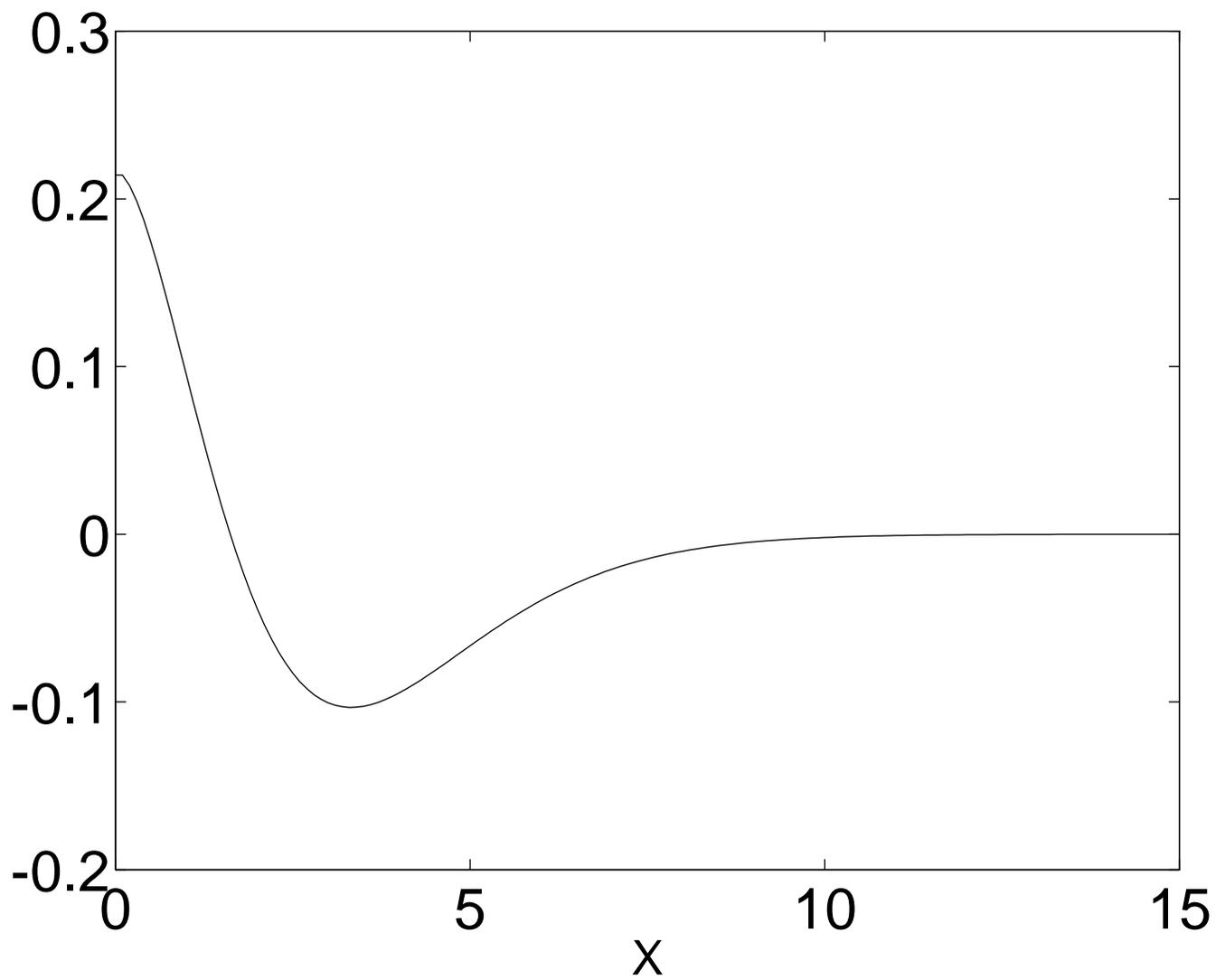
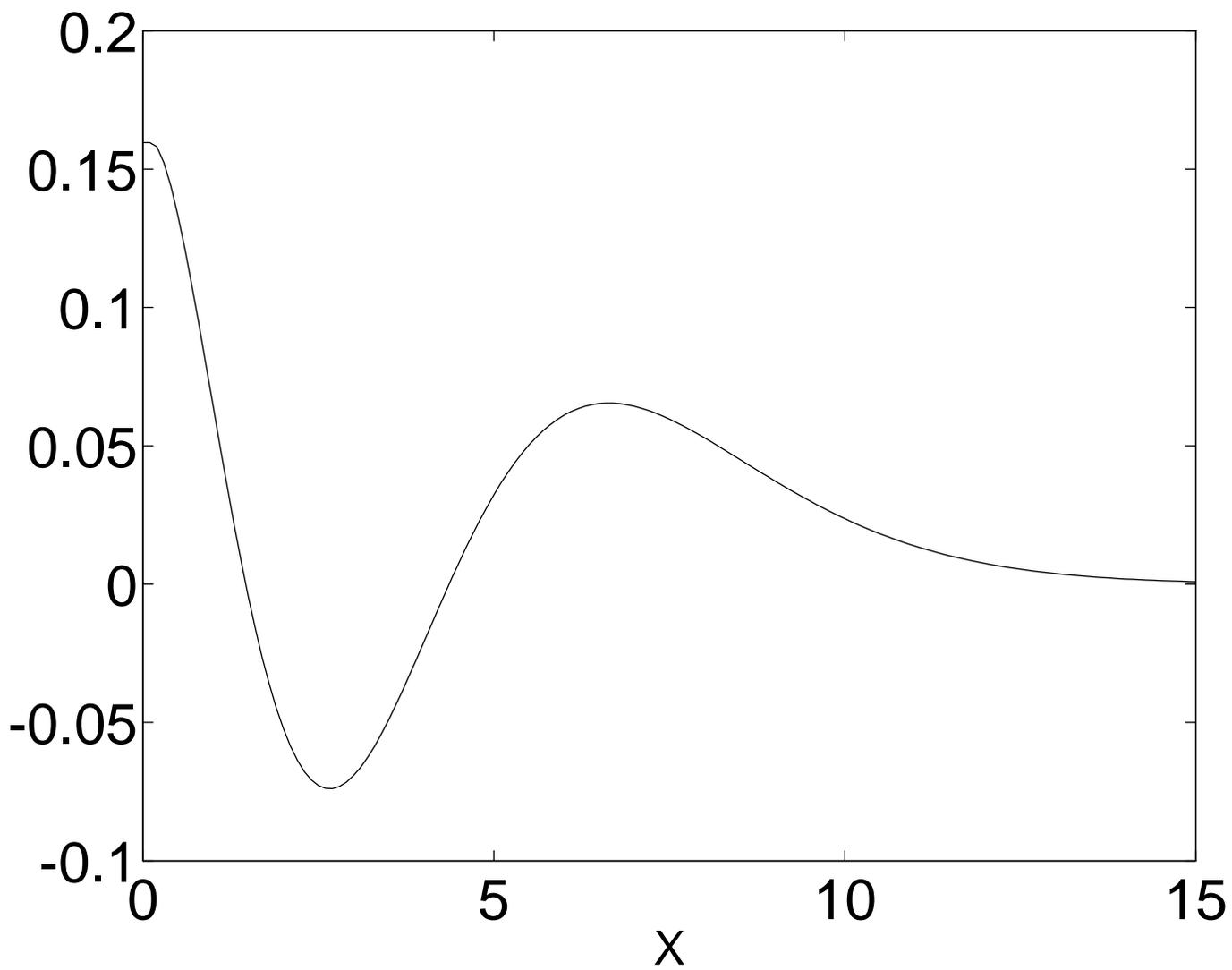
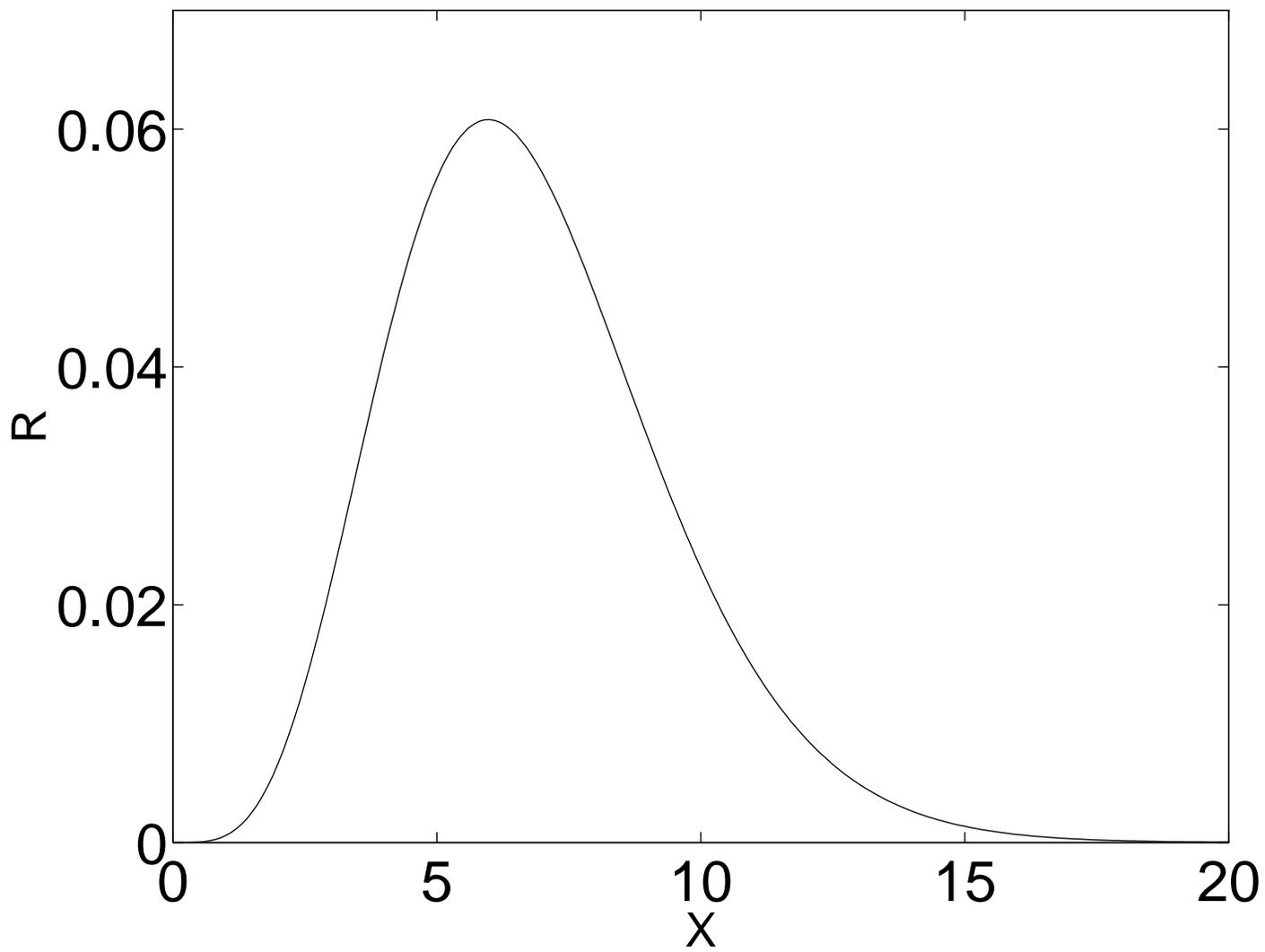
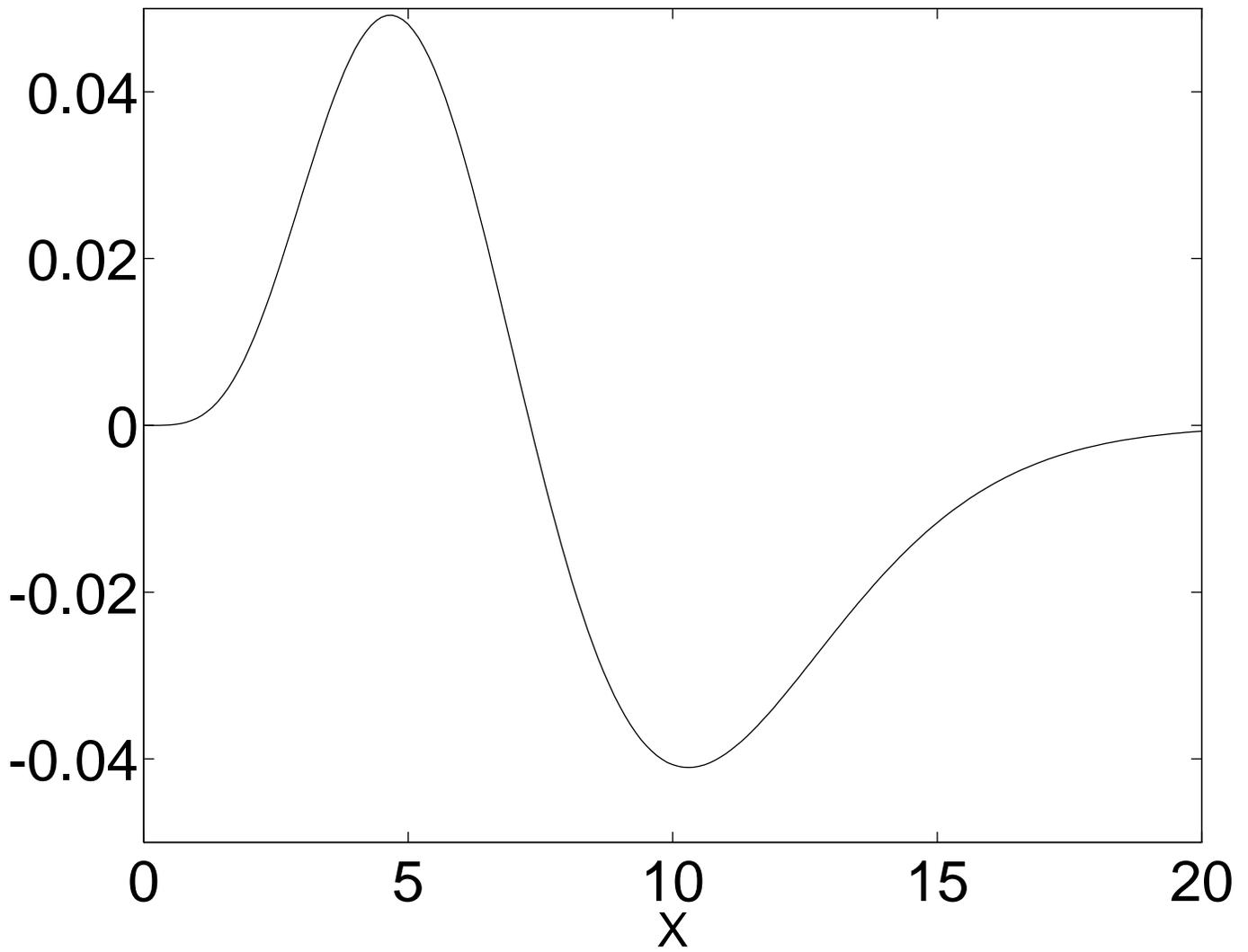
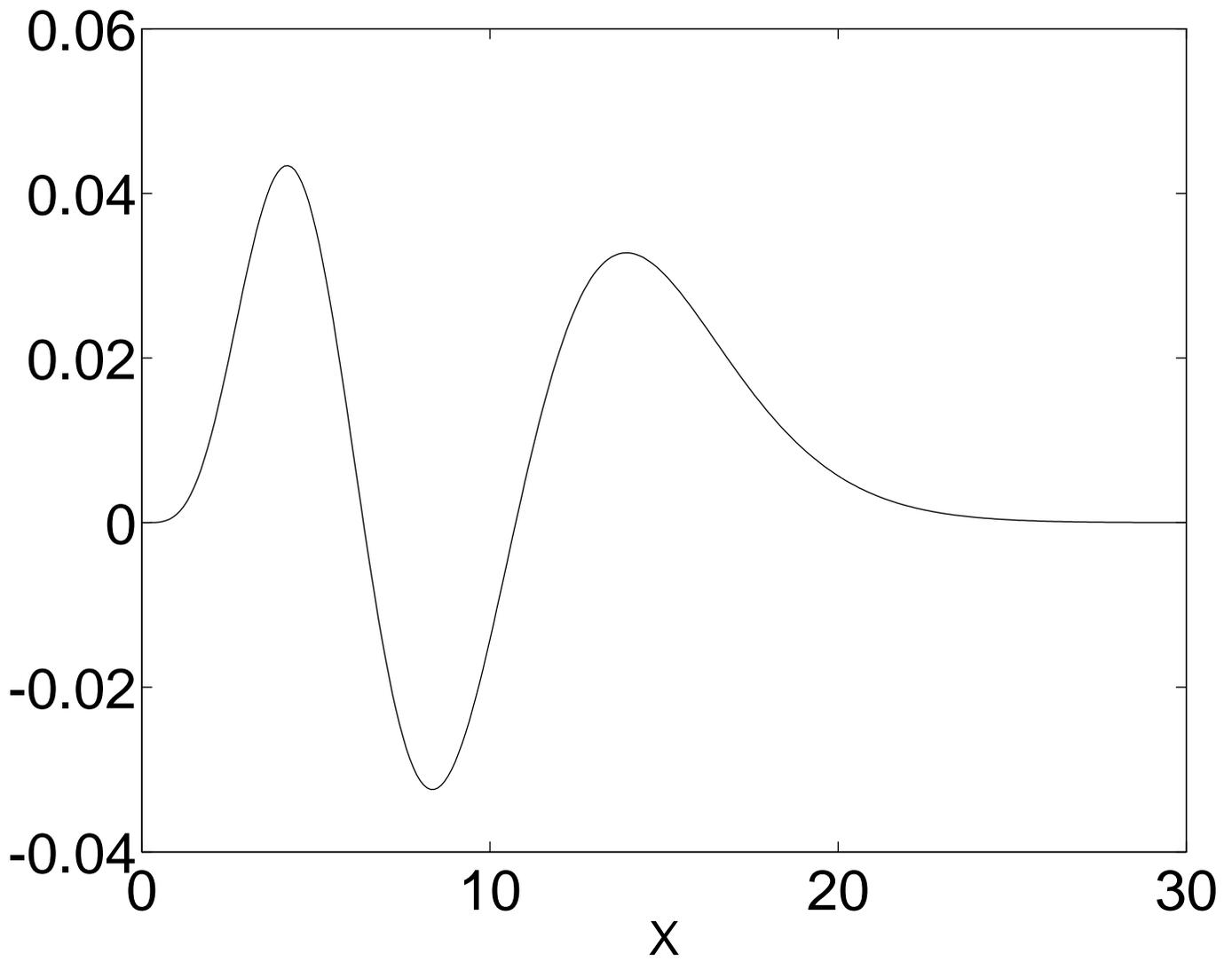
\end{document}